\documentclass[twocolumn,english,aps,prd,10]{revtex4-1}
\usepackage{bm}
\usepackage{amsfonts}
\usepackage{amssymb}
\usepackage{amsmath}
\usepackage{graphicx}

\begin{document}

\title{Time evolution of the QED vacuum in a uniform electric field:\\
Complete analytic solution by spinorial decomposition}

\author{Iwo Bialynicki-Birula}\email{birula@cft.edu.pl}
\author{\L{}ukasz Rudnicki}\email{rudnicki@cft.edu.pl}
\affiliation{Center for Theoretical Physics, Polish Academy of Sciences\\
Al. Lotnik\'ow 32/46, 02-668 Warsaw, Poland}

\begin{abstract}
Exact analytical solutions are presented for the time evolution of the density of pairs produced in the QED vacuum by a time-independent, uniform electric field. The mathematical tool used here to describe the pair production is the Dirac-Heisenberg-Wigner function introduced before [Phys. Rev. D {\bf 44}, 1825 (1991)]. The initial value problem for this function is solved by decomposing the solution into a product of spinors. The equations for spinors are much simpler and are solved analytically. These calculations are nonperturbative since pair production is due to quantum-mechanical tunneling and the explicit solutions clearly exhibit their nonanalytic behavior.
\end{abstract}
\maketitle

\section{Introduction}

The problem of pair creation from the QED vacuum by electromagnetic fields has been studied by various methods since the early days of quantum electrodynamics. The application of the one-particle Dirac equation leads to difficulties (Klein paradox) because pair creation is obviously a many-body problem. All paradoxes are resolved when this process is described by quantum field theory. Since pair creation by an external electromagnetic field involves quantum-mechanical tunneling, the process cannot be described in the framework of perturbation theory. Some time ago we introduced in \cite{bgr} a formalism based on a generalization of the Wigner function to quantum field theory. This allowed us to describe pair creation without resorting to perturbation theory. The basic tool in this formalism is the 16-component Dirac-Heisenberg-Wigner (DHW) function. The formalism based on the DHW function was further developed in \cite{sbr,bdr,sr,sr1} to yield quantum transport theory for relativistic spinning particles. In several papers \cite{cz,vge,eh,zh1,zh2,zh3} a relativistic generalization of the Wigner function was introduced. However, this approach is not suited for the solution of the initial value problem which is the main subject of this work. Other methods were also developed to treat the problem of pair production \cite{mottola,page,kleinert} but they have not led to an exact analytic solution. Such a solution is especially useful since the dependence on the field strength of the pair density is described by a function with an essential singularity.

In the general case of space-dependent fields the evolution equations for the components of the DHW function form a set of complicated integro-differential equations. However, for a time-independent, uniform electric field these equations become tractable since, as shown in \cite{bgr}, they can be reduced to the following three equations for the relevant components of the DHW function (notation is explained in the next section):
\begin{align}\label{orig}
&\left(\partial_\tau+\mathcal{E}\partial_q\right)\left[\begin{array}{c}
a_0(q,\tau)\\a_1(q,\tau)\\a_2(q,\tau)\end{array}\right]\nonumber\\
&=\left[\begin{array}{ccc}
0&\mathcal{E}/E_q^2&0\\-\mathcal{E}/E_q^2&0&-2E_q\\0&2E_q&0\end{array}\right]\left[\begin{array}{c}
a_0(q,\tau)\\a_1(q,\tau)\\a_2(q,\tau)\end{array}\right].
\end{align}
These equations were integrated numerically in \cite{bgr} and the Schwinger formula \cite{js} for pair creation was obtained. Recently, by an elaborate procedure a single solution of equations related to Eqs.~(\ref{orig}) was constructed as a combination of parabolic cylinder functions \cite{hag}. This discovery stimulated our return to the study of pair production by a uniform electric field.

In the present paper we construct not one but a complete set of solutions of Eqs.~(\ref{orig}). With the use of these solutions we solve the initial value problem for these equations with arbitrary initial conditions. The initial conditions that interest us most are those of the QED vacuum and we describe the time evolution in this case. Our solution clarifies the mechanism responsible for pair production and may explain some details of the vacuum decay studied recently \cite{lr}.

In Sec.~\ref{secevol} we recall basic definitions and we derive a simplified form of the evolution equations (\ref{orig}) best suited for our analysis. In Sec.~\ref{secprec} we exhibit the underlying $O(3)$ symmetry that enabled us to express the solutions of the evolution equations in terms of spinors. The method of spinorial decomposition introduced in Sec.~\ref{secspin} to lower the order of the differential equations by reducing the number of functions is quite general and might be applicable to a class of similar problems. Finally, in Sec.~\ref{fin} we describe the time evolution of the QED vacuum and we give physical interpretation of our results.

\section{The DHW function and its evolution}\label{secevol}

The Dirac-Heisenberg-Wigner function $W\left(\bm{r},\bm{p},t\right)$ for the quantized Dirac field in an external electromagnetic field was defined in \cite{bgr} as follows:
\begin{align}\label{wig}
W_{\alpha\beta}&(\bm{r},\bm{p},t)=-\frac{1}{2}\int d^3\!s\,\exp\left(-i\bm{p}\!\cdot\!\bm{s}/\hbar\right)e^{-i\varphi}\nonumber \\
&\left\langle\Phi\left|\left[\hat{\Psi}_{\alpha}
\left(\bm{r}+\bm{s}/2,t\right),
\hat{\Psi}_{\beta}^{\dagger}
\left(\bm{r}-\bm{s}/2,t\right)\right]\right|\Phi\right\rangle,
\end{align}
where the square bracket denotes the commutator of the quantized Dirac field operator $\mathbf{\hat{\Psi}}$ with its hermitian conjugate and the phase $\varphi$ is the line integral of the vector potential
\begin{equation}
\varphi=\frac{e}{\hbar}\int_{-1/2}^{1/2}d\lambda\,\bm{s}\!\cdot\!\bm{A}
\left(\bm{r}+\lambda\bm{s},t\right).\label{d7}
\end{equation}
The commutator was introduced to make the DHW function invariant under charge conjugation. The phase factor ensures its gauge invariance and, as a result, the momentum variable $\bm{p}$ is the kinetic momentum and not the gauge-dependent canonical momentum. The DHW matrix $W_{\alpha\beta}$ will be expanded into a set of Dirac matrices, as was done in \cite{bgr}
\begin{align}\label{wig1}
W\left(\bm{r},\bm{p},t\right)= \frac{1}{4}\left(\mathtt{I_{4}}f_{0}\left(\bm{r},\bm{p},t\right)
+\sum_{j=1}^{3}\rho_{j}f_{j}\left(\bm{r},\bm{p},t\right)\right.\nonumber\\
+\left.\bm{\sigma}\!\cdot\!\bm{g_{0}}\left(\bm{r},\bm{p},t\right)
+\sum_{j=1}^{3}\rho_{j}\bm{\sigma}\!\cdot\!\bm{g_{j}}
\left(\bm{r},\bm{p},t\right)\right).
\end{align}
Since the DHW matrix is hermitian, the expansion coefficients are all real functions. The Dirac equations for the field operators lead to evolution equations for the DHW function $W$ that can be rewritten as 16 equations for the functions $f_{i}$ and $\bm{g_{i}}$ ($i=0,\ldots,3$).

For arbitrary electric $\bm{E}\left(\bm{r},t\right)$ and magnetic $\bm{B}\left(\bm{r},t\right)$ fields the evolution equations were derived in \cite{bgr}. Here we shall need only a simple version of these equations valid for a time-independent, uniform electric field. Since we will be concerned only with the time evolution of the initial vacuum state, it is sufficient to consider only ten, out of 16, components of the DHW function that couple to the vacuum DHW function. The relevant components can be combined into a ten-dimensional vector $W=\left(f_3,\bm{g_0},\bm{g_1},\bm{g_2}\right)$ and the evolution equations have the form:
\begin{align}\label{eqm}
\left(\partial_t+e\bm{E}\!\cdot\!{\bm\partial}_{\bm{p}}\right)W\left(\bm{p},t\right)
=\frac{2c}{\hbar} {\hat M}\left(\bm{p}\right) W\left(\bm{p},t\right),
\end{align}
where we have taken into account that a uniform electric field does not introduce any space dependence. The $10\times10$ matrix ${\hat M}\left(\bm{p}\right)$ has the following block structure
\begin{displaymath}
{\hat M}\left(\bm{p}\right)=\left[
\begin{array}{c|ccc|ccc|ccc}
 0 & ~ & \bm{0} & ~ & ~ & \bm{0} & ~ & ~ & \bm{p} & ~\\
 \hline
 ~ & ~& ~& ~ & ~ & ~ & ~ & ~ & ~ & ~\\
 \bm{0} & ~& [\,0\,]_3 & ~ & ~ & {\bm p}\times & ~ & ~ & [\,0\,]_3 & ~\\
 ~ & ~ & ~ & ~ & ~ & ~ & ~ & ~ & ~ & ~\\
 \hline
 ~ & ~ & ~ & ~ & ~ & ~ & ~ & ~ & ~ & ~\\
 \bm{0} & ~ & {\bm p}\times & ~ & ~ & [\,0\,]_3 & ~ & ~ &-[\,mc\,]_3 & ~\\
 ~ &~ & ~ & ~ & ~ & ~ & ~ & ~ & ~ & ~ \\
 \hline
 ~ & ~ & ~ & ~ & ~ &~ & ~ & ~ & ~ & ~\\
 -\bm{p} & ~ & [\,0\,]_3 & ~ & ~ & [\,mc\,]_3 & ~ & ~ & [\,0\,]_3 & ~ \\
 ~ & ~ & ~ & ~ & ~ & ~ & ~ & ~ & ~ & ~
\end{array}
\right],
\end{displaymath}
where we introduced the following shorthand notation for the $3\times3$ matrices:
\begin{align}\label{shnot}
[\,0\,]_3=&\left[\begin{array}{ccc}0&0&0\\0&0&0\\0&0&0
\end{array}\right],\quad
[\,mc\,]_3=\left[\begin{array}{ccc}mc&0&0\\0&mc&0\\0&0&mc
\end{array}\right],\nonumber\\
&{\bm p}\times=\left[\begin{array}{ccc}0&p_z&-p_y\\-p_z&0&p_x\\p_y&-p_x&0
\end{array}\right].
\end{align}
>From now on we shall use units $c=1=\hbar$.

The vacuum DHW function contains only the components $f_3=-2m/E_p$ and $\bm{g_1}=-2\bm{p}/E_p$, where $E_p=\sqrt{m^2+\bm{p}^2}$. The solution of the evolution equations that begins as the vacuum state, as shown in \cite{bgr}, lies in a three-dimensional subspace and it can be described by the three-dimensional vector  $\left(w_1(\bm{p},t),w_2(\bm{p},t),w_3(\bm{p},t)\right)$,
\begin{align}\label{wu}
W\left(\bm{p},t\right)=-2\left[w_1(\bm{p},t)u_1+w_2(\bm{p},t)u_2+w_3(\bm{p},t)u_3\right],
\end{align}
where the ten-dimensional orthonormal vectors $u_k$, in the same block convention as used for ${\hat M}(\bm p)$, have the form
\begin{subequations}
\begin{align}\label{u}
u_1&=\frac{1}{m_\perp}\left[\;m\;|\;\bm{0}\;|\,{\bm p}_\perp\,|\;\bm{0}\;\right],\\
u_2&=\frac{1}{m_\perp}\left[\;0\;|\,{\bm p}_\perp\times{\bm n}\,|\;\bm{0}\;|\;m{\bm n}\;\right],\\
u_3&=\left[\;0\;|\;\bm{0}\;|\;\bm{n}\;|\;{\bm 0}\;\right],
\end{align}
\end{subequations}
where $m_\perp=\sqrt{m^2+p_\perp^2}$ is the transverse mass (energy), $\bm{n}$ is the unit vector in the field direction, and bold zeros stand for three-dimensional zero vectors. Note that the derivative in the field direction present in (\ref{eqm}) does not affect the vectors $u_k$ since they depend only on the transverse momentum ${\bm p}_\perp$. The DHW function of the vacuum state in this basis is
\begin{align}\label{vac}
W_{\rm vac}=-\frac{2m_\perp}{E_p}u_1-\frac{2p_\parallel}{E_p}u_3,
\end{align}
where $p_\parallel={\bm n}\!\cdot\!{\bm p}$. The subspace spanned by the vectors $u_k$ is invariant under the action of the matrix ${\hat M}(\bm p)$, namely
\begin{subequations}
\begin{align}
{\hat M}(\bm p)u_1&=-p_\parallel u_2,\\
{\hat M}(\bm p)u_2&=p_\parallel u_1-m_\perp u_3,\\
{\hat M}(\bm p)u_3&=m_\perp u_2.
\end{align}
\end{subequations}
With the use of these relations, by comparing the coefficients in Eq.~(\ref{eqm}) that multiply the vectors $u_k$, we obtain the following three equations for the coefficient functions
\begin{align}\label{eqmw}
\left(\partial_\tau+\mathcal{E}\partial_q\right)
\left[\begin{array}{c}w_1\\w_2\\w_3\end{array}\right]
=2\left[\begin{array}{ccc}0&q&0\\-q&0&1\\0&-1&0\end{array}\right]
\left[\begin{array}{c}w_1\\w_2\\w_3\end{array}\right],
\end{align}
where we introduced the following dimensionless variables ($c$ and $\hbar$ are reintroduced below for clarity)
\begin{align}
\tau=\frac{m_\perp c^2t}{\hbar},\quad q=\frac{p_\parallel}{m_\perp c},\quad
\mathcal{E}=\frac{e\hbar E}{m_\perp^2c^3}.
\end{align}
These evolution equations are equivalent to the original ones (\ref{orig}), treated in \cite{bgr} but their new form is more convenient. The new functions $(w_1,w_2,w_3)$ are related to the old ones $(a_0,a_1,a_2)$ by a rotation,
\begin{align}\label{rot}
w_1=\frac{a_0-qa_1}{\sqrt{1+q^2}},\quad w_2=a_2,\quad w_3=\frac{qa_0+a_1}{\sqrt{1+q^2}}.
\end{align}
The solution of the evolution equations (\ref{eqmw}) will be reduced in the next section to the solution of ordinary differential equations.

\section{Reduction to a precession equation}\label{secprec}

The solution of the initial value problem for the equations (\ref{eqmw}) will be reduced now to a set of ordinary differential equations by an extension of the method of characteristics. To this end let us denote by $\bm{\mathcal{X}}_k(q)$ three linearly independent vectors that are solutions of the following set of equations
\begin{align}\label{eqq}
\mathcal{E}\frac{d}{dq}\bm{\mathcal{X}}_k(q)={\hat L}(q){\bm{\mathcal{X}}}_k(q),
\end{align}
where ${\hat L}(q)$ is the matrix appearing in Eq.~(\ref{eqmw}) and $k$ labeling different solutions runs from 1 to 3. From the vectors ${\bm{\mathcal{X}}}_k(q)$ we can build a nonsingular matrix $\hat{\mathcal{W}}(q)$,
\begin{align}\label{matw}
\hat{\mathcal{W}}(q)=\left[\begin{array}{ccc}{\mathcal{X}}_1^1&{\mathcal{X}}_2^1&{\mathcal{X}}_3^1\\
{\mathcal{X}}_1^2&{\mathcal{X}}_2^2&{\mathcal{X}}_3^2\\
{\mathcal{X}}_1^3&{\mathcal{X}}_2^3&{\mathcal{X}}_3^3
\end{array}\right],
\end{align}
where the lower index labels different vectors and the upper index labels the vector components. The solution of the initial value problem for Eq.~(\ref{eqmw}) has the form
\begin{align}\label{init}
{\bm{\mathcal{X}}}(q,\tau)=\hat{\mathcal{W}}(q)
\left[\hat{\mathcal{W}}(q-\mathcal{E}\tau)\right]^{-1}
{\bm{\mathcal{X}}}(q-\mathcal{E}\tau,0).
\end{align}
The validity of this formula can be checked by inspection. First, we check that this expression is a solution of Eq.~(\ref{eqmw}). The combination of partial derivatives $\left(\partial_\tau+\mathcal{E}\partial_q\right)$ when acting on the first part of (\ref{init}) reduces to the derivative with respect to $q$ while it does not affect the part dependent on $q-\mathcal{E}\tau$. The derivative with respect to $q$ reproduces the right hand side of equations. Second, we check that this expression satisfies the initial condition. This follows from the fact that at $\tau=0$ both arguments become equal and the product of the matrix and its inverse reduces to the unit matrix.

To make use of the formula (\ref{init}) we must find a complete set of solutions of Eq.~(\ref{eqq}). The reduction of the set of three equations (\ref{eqq}) to just one equation for a single function leads to a third-order equation. In contrast to second-order equations, such equations are, in general, hard to solve. A representation of the solution of the resulting third-order equation in the form of a contour integral has been given by Rado\.zycki \cite{torado}. The contour integral form was sufficient to obtain the Schwinger formula. However, it cannot serve as a basis for the solution of the initial value problem and we proceed by a different route.

The key property is the presence of the $O(3)$ symmetry in the equation (\ref{eqq}). Indeed, this is an equation describing the precession in three-dimensional space of the vector ${\bm{\mathcal{X}}}(q)$ around the vector ${\bm{\mathcal{H}}}(q)$,
\begin{align}\label{prec}
\mathcal{E}{{\bm{\mathcal{X}}}'}(q)=-2\,{\bm{\mathcal{H}}}(q)\times{\bm{\mathcal{X}}}(q),
\end{align}
where ${\bm{\mathcal{H}}}=\left(1,0,q\right)$ and $q$ plays the role of time. The factor of 2 is pulled out for further convenience. Were it not for the $q$-dependence of the precession vector ${\bm{\mathcal{H}}}(q)$, the solution of Eq.~(\ref{prec}) would have been very simple. Still, even for an arbitrary ${\bm{\mathcal{H}}}(q)$ the precession equations have some special properties implied by the $O(3)$ symmetry that will enable us to find their solutions. In the next section we proceed directly to the solution of the equations (\ref{prec}). In the Appendix we explain the group-theoretic foundations of this solution.

\section{Solution of the precession equations by spinorial decomposition}\label{secspin}

In order to find analytic solutions of the precession equations, we express the vector ${\bm{\mathcal{X}}}(q)$ as a composition of a two-component spinor $\Psi(q)$ and its complex conjugate,
\begin{align}\label{comp}
{\bm{\mathcal{X}}}(q)=\Psi^\dagger(q)\bm{\sigma}\Psi(q),
\end{align}
where $\bm{\sigma}$ is the vector of Pauli matrices. The precession equation (\ref{prec}) will be satisfied by the vector ${\bm{\mathcal{X}}}(q)$ if the spinor $\Psi(q)$ obeys the equation
\begin{align}\label{seq}
\mathcal{E}\Psi'(q)=i\left[\bm{\sigma}\!\cdot\!{\bm{\mathcal{H}}}(q)\right]\Psi(q).
\end{align}
This set of two coupled equations can always be reduced to separate second-order equations for the two components of the spinor. This is a crucial simplification--- equations of the second-order are much more friendly than third-order equations. The technique of lowering the order of a differential equation by seeking the solution in the product form was known before for some third-order equations \cite{ww}. Here we extend it to a set of equations with an underlying symmetry group.

For our choice of the precession vector we obtain from (\ref{seq}),
\begin{align}\label{pcyl}
\mathcal{E}^2\psi''(q)+\left(1+q^2\mp i\mathcal{E}\right)\psi(q)=0,
\end{align}
where the upper (lower) sign corresponds to the upper (lower) component $\psi(q)$ of the spinor $\Psi(q)$. These equations have parabolic cylinder functions as their solutions. Of course, the solutions of the second-order equations must be chosen in such a way that the spinor $\Psi(q)$ is a solution of the original equations (\ref{seq}). There are several equivalent representations of the solutions of Eqs.~(\ref{pcyl}) in terms of parabolic cylinder functions (cf., for example, \cite{nist}). We have chosen the set of two orthonormal spinors in such a way that they contain only two parabolic cylinder functions and their complex conjugates,
\begin{align}\label{two}
\Psi_1=\left[\begin{array}{c}F(\mathcal{E},q)\\-G(\mathcal{E},q))
\end{array}\right],\quad
\Psi_2=\left[\begin{array}{c}G^*(\mathcal{E},q)\\F^*(\mathcal{E},q))
\end{array}\right],
\end{align}
where
\begin{subequations}\label{pcf}
\begin{align}
F(\mathcal{E},q)&=e^{-\frac{\pi}{8\mathcal{E}}}
\frac{1}{\sqrt{2\mathcal{E}}}
D_{-1-\frac{i}{2\mathcal{E}}}\left(\frac{1+i}{\sqrt{\mathcal{E}}}q\right),\\
G(\mathcal{E},q)&=e^{-\frac{\pi}{8\mathcal{E}}}\frac{1-i}{\sqrt{2}}
D_{-\frac{i}{2\mathcal{E}}}\left(\frac{1+i}{\sqrt{\mathcal{E}}}q\right).
\end{align}
\end{subequations}
The two complex spinors (\ref{two}) will serve as the building blocks in the construction of three real orthonormal vectors. We shall represent them in a compact way by employing the same Dirac matrices (although in a different context) as in the formula (\ref{wig1}),
\begin{align}\label{three}
{\bm{\mathcal{X}}}_k=-\frac{1}{2}\Phi^\dagger\rho_k\bm{\sigma}\Phi,
\end{align}
where
\begin{align}\label{phi}
\Phi=\left[\begin{array}{c}\Psi_1\\\Psi_2\end{array}\right],
\end{align}
and the matrices $\rho_k$ act in the space spanned by spinors $\Psi_1$ and $\Psi_2$. For example, ${\bm{\mathcal{X}}}_1=-(\Psi^\dagger_1{\bm\sigma}\Psi_2+\Psi^\dagger_2{\bm\sigma}\Psi_1)/2$. Having determined an orthonormal set of vectors that solve the precession equations (\ref{prec}), we may construct the matrix ${\hat{\mathcal{W}}}$. It is made of various quadratic expressions built of the two functions $F$ and $G$
\begin{align}\label{3in2}
{\hat{\mathcal{W}}}=\left[\begin{array}{ccc}\Re[F^2-G^2]&\Im[F^2-G^2]&2\Re[FG^*]\\
\Im[F^2+G^2]&\Re[F^2+G^2]&-2\Im[FG^*]\\
-2\Re[FG]&2\Im[FG]&-FF^*+GG^*\end{array}\right].
\end{align}
One may check by a direct calculation that this matrix is orthonormal ${\hat{\mathcal{W}}}^T{\hat{\mathcal{W}}}={\bm{\mathrm 1}}$. Owing to the minus sign in the definition (\ref{three}), the determinant of ${\hat{\mathcal{W}}}$ is equal to 1 (it describes a pure rotation). The final formula for the vector ${\bm{\mathcal{X}}}(q,\tau)$ that represents the solution of the initial value problem is
\begin{align}\label{initf}
{\bm{\mathcal{X}}}(q,\tau)=\hat{\mathcal{W}}(q)\!\cdot\!
\left[\hat{\mathcal{W}}(q-\mathcal{E}\tau)\right]^T
\!\cdot\!{\bm{\mathcal{X}}}(q-\mathcal{E}\tau,0),
\end{align}
where the matrices ${\hat{\mathcal{W}}}$ are expressed in terms of the parabolic cylinder functions $F$ and $G$ by the formula (\ref{3in2}). Note that as compared with the general formula (\ref{init}) we could replace the inverse matrix with the transposed matrix because we used an orthonormal set of solutions.

\section{Time evolution of the QED vacuum}\label{fin}

The solution of the initial value problem (\ref{initf}) can be used to evaluate the overlap $Q(q,\tau)$ between the DHW function evolved from the QED vacuum in time and the initial DHW function. This overlap is a simple measure of the departure from the initial state. It is given by the projection of the vector (\ref{initf}) on the vector
\begin{align}\label{vv}
{\bm{V}_0}(q)=\frac{1}{\sqrt{1+q^2}}\left(1,0,q\right),
\end{align}
that describes the initial vacuum state. In this way we obtain the following compact formula for the overlap as a scalar product of two vectors,
\begin{align}\label{qvv}
Q(q,\tau)={\bm V}(q)\!\cdot\!{\bm V}(q-\mathcal{E}\tau),
\end{align}
where
\begin{align}\label{vq}
{\bm V}(q)=\hat{\mathcal{W}}^T(q)\!\cdot\!{\bm{V}_0}(q).
\end{align}
\begin{figure}
\centering
\includegraphics[scale=0.5]{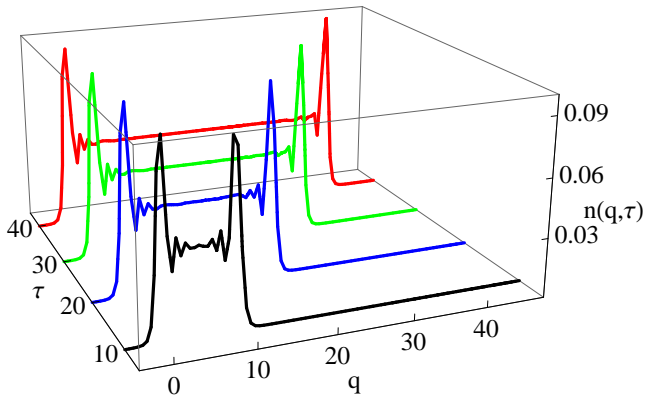}
\caption{(Color online) The pair density $n(q,\tau)$ as a function of the dimensionless momentum $q$ plotted at the values 10, 20, 30, and 40 of the dimensionless time $\tau$. The dimensionless field strength $\mathcal{E}$ is set to one.}\label{fig1}
\end{figure}

The overlap function, denoted by $a_0$ in \cite{bgr}, was obtained there by numerical integration of Eqs.~(\ref{orig}). Our analytic formulas reproduce exactly those results but instead of showing again the plots we shall transform the overlap function into a more intuitive characteristic --- the density of pairs $n(q,\tau)$ produced by the electric field. The relation between these two functions, as shown in \cite{bgr} by the calculation of the produced current, is very simple
\begin{align}\label{n}
n(q,\tau)=\frac{1-Q(q,\tau)}{2}.
\end{align}
Obviously, the number of pairs is small when the overlap function is close to unity. In Fig.~\ref{fig1} we show the pair density as a function of $q$ for different values of $\tau$. The pair density $n(q,\tau)$ as a function of $q$ has a plateau in the middle region and some transient oscillations near $q\approx 0$ and $q\approx \mathcal{E}\tau$. The length of the plateau region grows with $\tau$ covering more and more of the $q$ axis. The plateau region reaches only as far as $\mathcal{E}\tau$. This is an expression of causality because the largest momentum that could be acquired in time $\tau$ is $\mathcal{E}\tau$. The departures at $q=0$ and $q\approx \mathcal{E}\tau$ from the plateau values can be explained as transient effects. The longer we apply the electric field, the less important is their role: they occupy a smaller and smaller  fraction of the phase space.

In the limit, when $\tau\to\infty$, the pair density is totally dominated by the plateau. The height of the plateau at the mid point for $q=\mathcal{E}\tau/2$ has the value
\begin{align}\label{mid}
(1-Q(\mathcal{E}\tau/2,\tau))/2=(1-{\bm V}(\mathcal{E}\tau/2)\!\cdot\!{\bm V}(-\mathcal{E}\tau/2))/2.
\end{align}
This height can be determined as a function of $\mathcal{E}$, in the limit when $\tau\to\infty$, from the asymptotic expansions of the parabolic cylinder functions (\ref{pcf}) for large values of the argument. These expansions are different for $q\to\infty$ and $q\to-\infty$ due to the Stokes phenomenon. The role of the Stokes phenomenon in pair production has been recently exposed in the context of the WKB approximation \cite{stokes}. The leading terms in the asymptotic expansions are \cite{gr}
\begin{subequations}\label{as}
\begin{align}
F(\mathcal{E},|q|)&=O\left(\frac{1}{|q|}\right),\label{as1}\\
G(\mathcal{E},|q|)&=e^{-i\Phi_1(q)-i\pi/4}+O\left(\frac{1}{|q|}\right),\label{as2}\\
F(\mathcal{E},-|q|)&=\frac{e^{-\frac{\pi}{4\mathcal{E}}}
\sqrt{\frac{\pi}{\mathcal{E}}}e^{-i\Phi_2(q)}}
{\Gamma\left(1+\frac{i}{2\mathcal{E}}\right)}\!+O\left(\frac{1}{|q|}\right),\label{as3}\\
G(\mathcal{E},-|q|)&=e^{-\frac{\pi}{2\mathcal{E}}}e^{-i\Phi_1(q)-i\pi/4}+O\left(\frac{1}{|q|}\right)\label{as4},
\end{align}
\end{subequations}
where the phases $\Phi(q)$ are
\begin{align}\label{phase}
\Phi_1(q)=\frac{\ln(2q^2)-\ln(\mathcal{E})}{4\mathcal{E}},\quad \Phi_2(q)=\frac{\ln(2q^2)}{4\mathcal{E}}.
\end{align}
Now, let us note that for $|q|\gg1$ the vacuum vector (\ref{vv}) reduces to $(0,0,{\rm sgn}(q))$. Therefore, the vectors ${\bm V}(\pm\infty)$ involve only those elements of the matrix (\ref{3in2}) that form its bottom row. This observation leads to a particularly simple result for the positive values of $q$. In this case, according to (\ref{as1}) and (\ref{as2}), we obtain
\begin{align}\label{limp}
{\bm V}(\infty)=(0,0,1).
\end{align}
The calculation of ${\bm V}(-\infty)$ is more involved, since the asymptotic formulas are more complicated. However, because ${\bm V}(\infty)$ has only one nonvanishing component, we only need the last component of the vector ${\bm V}(-\infty)$ to calculate the scalar product (\ref{qvv}). This component is equal to $|F|^2-|G|^2$, where we can substitute the asymptotic values given by (\ref{as3}) and (\ref{as4}). With the use of the following identity for the Euler $\Gamma$ function:
\begin{align}\label{gamma}
\Gamma\left(1+ix\right)\Gamma\left(1-ix\right)=\frac{\pi x}{\sinh(\pi x)},
\end{align}
we obtain finally
\begin{align}\label{limn}
\lim_{\tau\to\infty}n(\mathcal{E}\tau/2,\tau)=\exp\left(-\frac{\pi}{\mathcal{E}}\right).
\end{align}
It was shown before (cf., Eqs.~(69), (72), and (73) in \cite{bgr}) that this value for the number of pairs, after integration over all momenta, leads to the Schwinger formula \cite{js}.

\section{Conclusions}

We have shown once more that the DHW function is a powerful tool to analyze quantum electrodynamic processes without recourse to perturbation theory. It is true that the evolution equations for this function, in general case, are prohibitively complicated. However, the case of the time-independent, uniform electric field treated here allows for the complete analytic solution of the initial value problem. Our derivation of the analytic solution may also be viewed as a case study to explain a new method of solving differential equations with some group symmetry. The original equations involved the adjoint representation of the $O(3)$ group. By expressing the solutions in terms of the fundamental representation (spinors) we were able to simplify the equations. This technique will also work for other symmetry groups.

\acknowledgments

We thank Marek Ku\'s and Adam Sawicki for discussions. This research was partly supported by the grant from the Polish Ministry of Science and Higher Education for the years 2010--2012.
\appendix
\section{}
In order to see the underlying group structure, we shall rewrite Eq.~(\ref{prec}) with the use of the structure constants $c^i_{jk}$ for the $O(3)$ group. To show the general structure of these equations more clearly, we will change the notation as follows:
\begin{align}\label{group}
\frac{d}{d\lambda}{\mathcal{G}}^i(\lambda)
=-c^i_{jk}{\mathcal{H}}^j(\lambda){\mathcal{G}}^k(\lambda).
\end{align}
The structure constants for the $O(3)$ group are given by the Levi-Civita symbols $\epsilon_{ijk}$.

There are two equivalent ways of looking at these equations. One may consider the right hand side as the Poisson bracket of ${\mathcal{H}}^j$ and ${\mathcal{G}}^k$. Then the equations determine a family of canonical transformations. Alternatively, one may view them as the evolution equation of a vector ${\bm{\mathcal{X}}}$ in the adjoint representation of the group, i.e.
\begin{align}\label{adj}
\frac{d}{d\lambda}{\bm{\mathcal{X}}}(\lambda)
=i\left[{\bm S}\!\cdot\!{\bm{\mathcal{H}}}(\lambda)\right]
{\bm{\mathcal{X}}}(\lambda),
\end{align}
where the matrices $S_i$ are the spin 1 matrices in the Cartesian basis. The second interpretation is closer to our procedure. We just replace the evolution equations (\ref{adj}) for the vector in the adjoint representation (spin 1) by the evolution equations for the vector in the fundamental representation (spin 1/2). We only have to change the generators ${\bm S}\to{\bm\sigma}/2$ but we keep the same combination of the generators as determined by ${\bm{\mathcal{H}}}$. After solving the equations in the spinorial representation we may construct solutions in any other representation by taking the appropriate products.

\end{document}